# Controlling synchronization in large laser networks using number theory

Micha Nixon[1], Moti Fridman[1], Eitan Ronen[1], Asher A. Friesem[1], Nir Davidson[1] & Ido Kanter[2]

[1]*Dept. of Physics of Complex Systems, Weizmann Institute of Science, Rehovot 76100, Israel.*

[2]*Dept. of Physics, Bar-Ilan University, Ramat-Gan 52900, Israel.*

**Synchronization in networks with delayed coupling are ubiquitous in nature and play a key role in almost all fields of science including physics, biology, ecology, climatology and sociology [1-7]. In general, the published works on network synchronization are based on data analysis and simulations, with little experimental verification [8,9,10]. Here we develop and experimentally demonstrate various multi-cluster phase synchronization scenarios within coupled laser networks. Synchronization is controlled by the network connectivity in accordance to number theory, whereby the number of synchronized clusters equals the greatest common divisor of network loops. This dependence enables remote switching mechanisms to control the optical phase coherence among distant lasers by local network connectivity adjustments. Our results serve as a benchmark for a broad range of coupled oscillators in science and technology, and offer feasible routes to achieve multi-user secure protocols in communication networks and parallel distribution of versatile complex combinatorial tasks in optical computers.**

Theoretical investigations suggest that the underlying properties that govern network dynamics can be attributed to either the *statistical properties* such as degree distribution [11,12] or the connectivity detailed properties such as *extreme eigenvalues* of the graph Laplacian [13,14]. Experimentally, the relation between the symmetry of small networks, and their synchronization states were observed [7,8,9,10]. Recently, the emergence of synchronized clusters were also observed in small laser networks with homogeneous delay times[9], where the number of clusters was limited to two as a result of the bidirectional coupling given by the time reversal symmetry of light.

Here we develop an approach for multi-cluster synchronization of larger networks of coupled lasers with homogeneous as well as with heterogeneous delay times. We exploit the Faraday effect to control the polarization degree of freedom in order to break the time reversal symmetry of light, resulting in unidirectional couplings and the formation of up to 16 clusters. The experimental arrangements and representative experimental results are presented in Fig. 1. The experimental arrangements include a degenerate laser cavity (Fig. 1a, centre) that can support many independent lasers [9,15], a coupling arrangement (Fig. 1a, right) for controlling the connectivities and obtaining unidirectional couplings between lasers and a detection arrangement (Fig. 1a, left) for detecting the far field (FF) intensity distributions from the lasers (see Methods). The insets (Fig. 1a, centre) show the rear near field (NF) intensity pattern of 8 beams corresponding to a mask of 8 holes and the front NF intensity pattern of 16 independent lasers (a pair of 8 lasers with orthogonal polarizations) which were obtained by using calcite beam displacer in the degenerate cavity. The phase independence among the uncoupled lasers is experimentally verified by comparing the far-fields (FF) intensity distributions of a single laser and 16 lasers (insets Fig. 1a, left). As evident, the lack of interference fringes and the essentially identical distributions indicate that there is no phase synchronization among the 16 lasers.

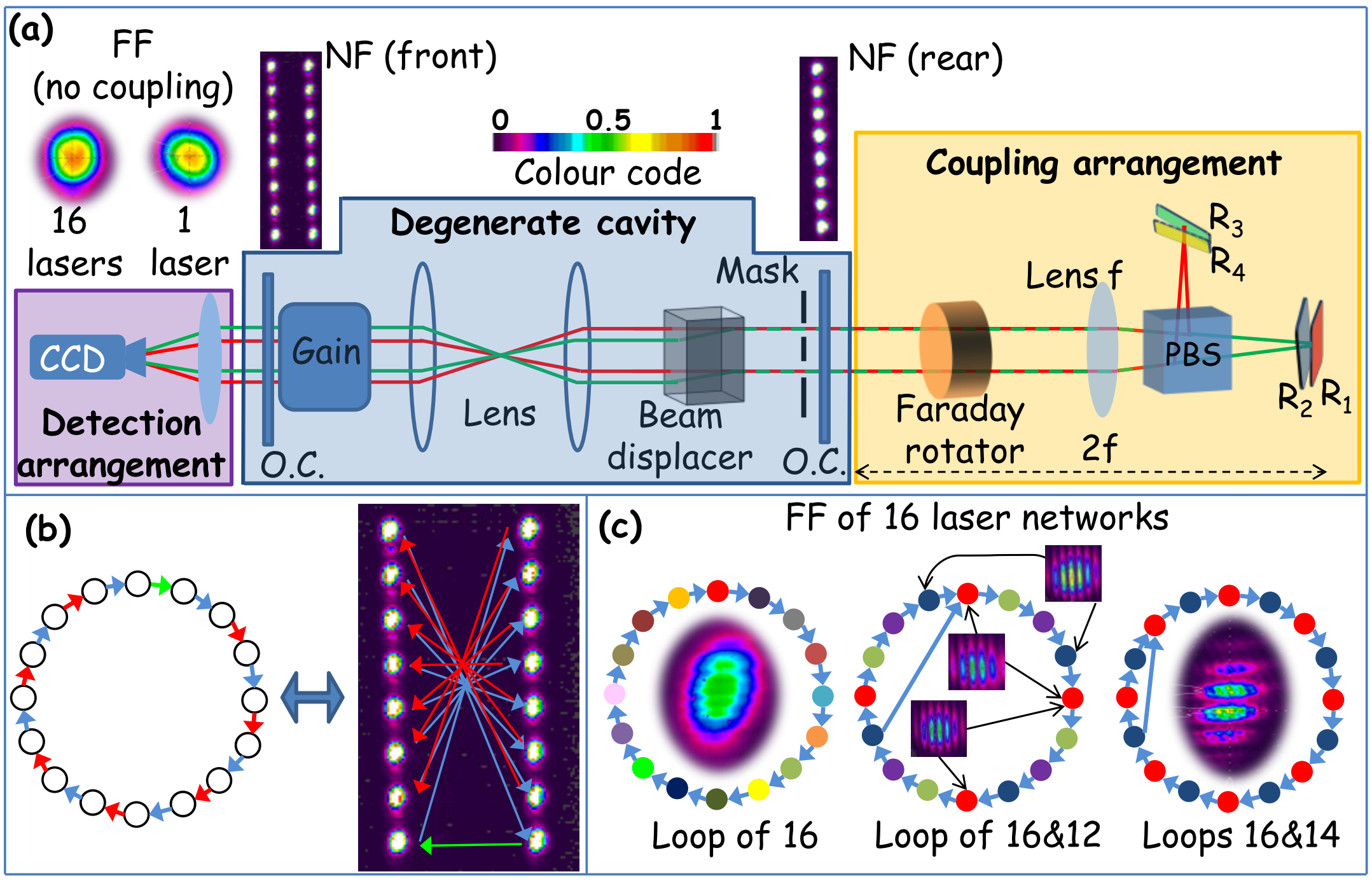

**Figure 1| Experimental arrangements and representative experimental results. a, A schematic sketch of a degenerate cavity that supports many independent uncoupled lasers as verified by the similarity between the detected far-fields (FF) intensity distributions of a single laser and 16 uncoupled lasers, a coupling arrangement with four mirrors for obtaining a variety of different coupling connectivities, and a detection arrangement for detecting the phase synchronization between any desired set of lasers with a CCD camera. b, Connectivity arrangement in a unidirectional loop of 16 lasers. The coloured arrows, added to the near-field (NF) intensity distributions of 16 lasers, denote which pairs of lasers are coupled by which mirrors. c, Three different networks of 16 lasers all with 4 *ns* unidirectional time-delayed couplings. For a single directed loop of 16 lasers, the FF intensity distribution indicates the lack of synchronization among all 16 lasers (left). For a network with 16 and 12 laser loops, the FF intensity distributions of different pairs of lasers indicates four synchronized clusters, each including lasers marked by a specific colour (centre). For a network with 16 and 14 laser loops the FF intensity distribution indicates two separate synchronized clusters (right).**

A variety of coupling connectivities can be realized by controlling the angular orientations of the four mirrors in the coupling arrangement. One is illustrated for a directed loop of 16 lasers (Fig. 1b), obtained by using three coupling mirrors (red, blue and green). The coloured arrows in the NF intensity pattern denote which one of the three mirrors led to a specific unidirectional coupling (Supplementary Information Figs. 1-3). We found that a directed loop of 16 coupled lasers does not lead to synchronization between any pair of lasers, as verified by the very poor fringe contrast in the FF intensity pattern of all 16 lasers (Fig. 1c, left). As the coupling delay time $\tau \approx 4$ *ns*, given by the round trip propagation time through the coupling arrangement ($\tau = 4f/c$), is much longer than the coherence time of the lasers $\tau_{coh} \approx 10$ *ps*, no synchronization is expected because the coupling signals arrives long after phase memory is lost. However, by using the fourth coupling mirror (yellow) to add a unidirectional coupling which forms a new directed loop of 14 lasers, synchronized cluster of alternating lasers emerges (Fig. 1c, right). The resulting FF intensity pattern with fringes only along the vertical direction indicates that synchronization now occurs only between lasers positioned along the same vertical column. Specifically, the network splits into two distinct synchronized clusters of lasers (denoted by either blue or red colours), i.e. all odd or all even lasers are synchronized, but pairs of odd-even lasers are not synchronized. Alternatively, the fourth coupling mirror could be used to couple between other pairs of lasers so as to form an additional loop of 12 lasers rather than 14 lasers (Fig. 1c, centre). Now, four distinct synchronized clusters emerge, as exemplified by the high contrast interference fringes of different pairs belonging to the same cluster.

The number of synchronized clusters can be predicted in accordance to the network connectivity by resorting to a simple relation that is based on number theory. Specifically, for homogeneous networks, the number of synchronized clusters is predicted to be equal to the greatest common divisor (GCD) of the network loops [16,17]. This is consistent with our experimental results, (Fig. 1c), where a network with 16 and 14 laser loops results in GCD(16,14)=2 synchronized clusters, a network of 16 and 12 laser loops results in GCD(16,12)=4 synchronized clusters, and a single directed loop of 16 lasers results in GCD(16)=16 synchronized clusters each comprised of a single laser.

The GCD rule for the number of clusters can be intuitively understood by the fact that each laser synchronizes to the delayed incoming signal and relays the optical phase information onwards in accordance to the network connectivity. As a result, in a directed loop of $n$ lasers with a coupling delay time of $\tau$, each laser is synchronized to its own signal delayed by $n{\cdot}\tau$. Consequently, the signal from each laser has $n{\cdot}\tau$ periodicity, however, no synchronized pairs of lasers exists, resulting in $n$ clusters. For a network with an additional loop of $m$ lasers, the signals from all lasers have to fulfil $n{\cdot}\tau$ and $m{\cdot}\tau$ periodicities, resulting in GCD($n$,$m$)$\cdot\tau$ periodicity and GCD($n$,$m$) synchronized clusters. Note that other periodic solutions that consist of fewer numbers of clusters also exist, but are unstable due to the information mixing mechanism15 [16,17].

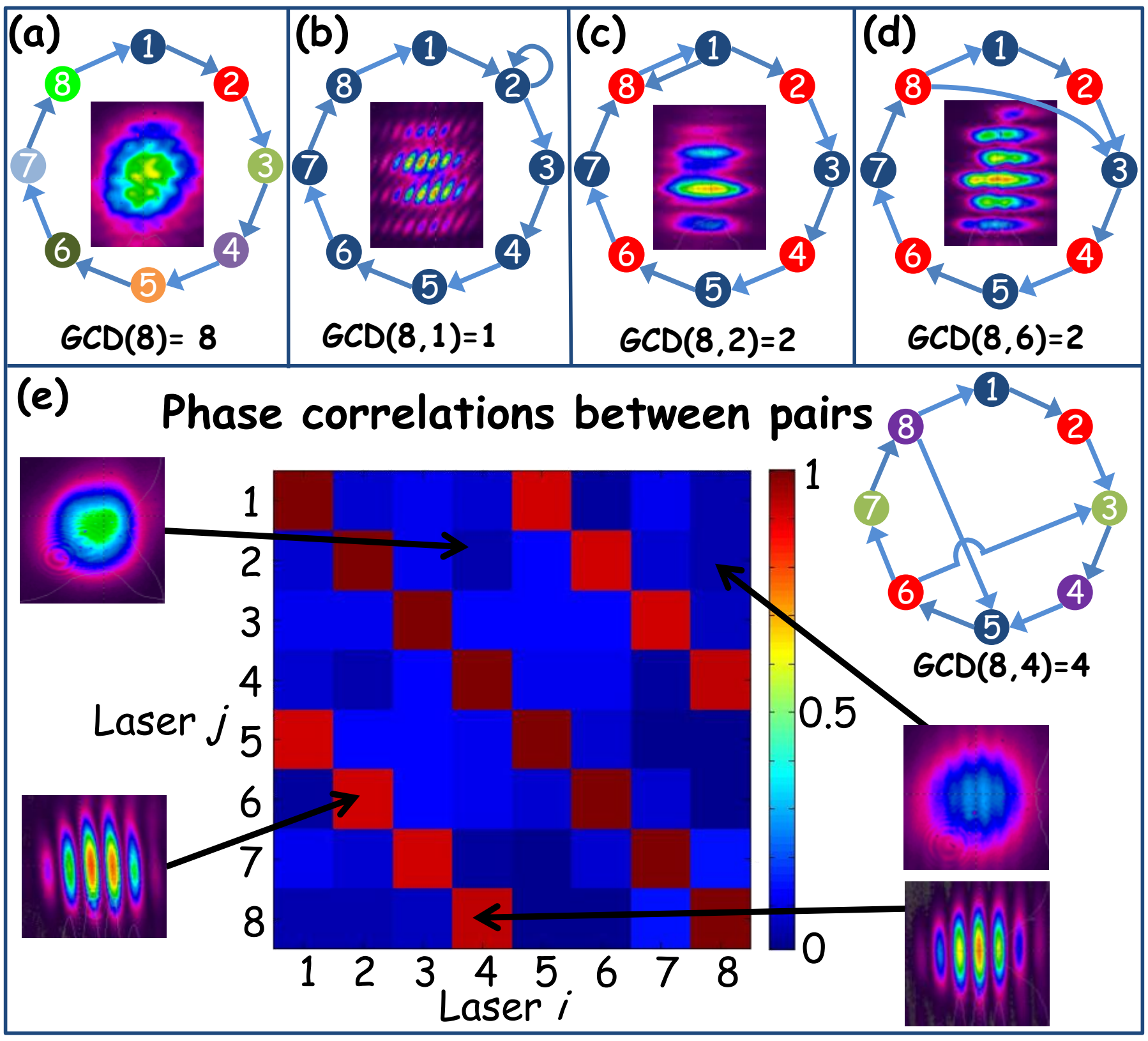

**Figure 2| Experimental phase synchronization in homogeneous networks. a, A directed loop of 8 lasers, where the lack of interference fringes in the FF intensity distribution of all 8 lasers, indicate no synchronization and eight separate clusters. b, The directed loop of 8 lasers with an additional self-feedback loop, where high contrast interference fringes along both horizontal and vertical directions, indicate high synchronization with one cluster GDC(8,1)=1. c, The directed loop of 8 lasers with an additional bidirectional loop of size 2, where high contrast interference fringes along the vertical direction only indicate two clusters, GDC(8,2)=2. d, The directed loop of eight lasers with additional loop of six lasers obtained with unidirectional coupling between laser 8 to laser 3, where high contrast interference fringes along the vertical direction only indicate two clusters, GDC(8,6)=2. e, The phase correlation (fringe visibility) between all pairs of lasers for a network with directed loops of 8 and 4 lasers, indicating four clusters GCD(8,4)=4.**

The experimental proof of concept of multi-user synchronization controlled by the GCD of homogeneous time delay networks is exemplified for various 8 laser networks (Fig. 2). A directed loop of 8 lasers exhibits no synchronization, whereby there are no interference fringes in the FF intensity pattern (Fig. 2a). A clear manifestation of non-local synchronization mechanism is exemplified in Fig. 2b where a single local network connectivity adjustment, remotely switches the synchronization state among all lasers. Specifically, the addition of a single self-feedback loop is sufficient to obtain a high degree of global synchronization among all lasers, forming a single synchronized cluster GCD(8,1)=1. The high contrast interference fringes appearing along both vertical and horizontal directions in the FF intensity pattern indicate that synchronization indeed occurs among all lasers in the network (Fig, 2b). A single bidirectional coupling channel added to the directed loop forms a loop of size 2 resulting in GCD(8,2)=2 synchronized clusters (Fig. 2c). The high contrast interference fringes appearing along the vertical axis only in the FF intensity pattern indicate that synchronization occurs only between lasers positioned along the same vertical column (Fig. 2c). Alternatively, two clusters can also be formed by adding unidirectional coupling from laser 8 to laser 3 to obtain a six laser loop and GCD(8,6)=2 clusters (Fig. 2d). A network consisting of 4 and 8 laser loops results in GCD(8,4)=4 clusters (Fig. 2e). This was quantified by measuring the second order phase correlation (fringe visibility[18]) between all possible 28 pairs of lasers (Fig. 2e, centre). Representative FF interference patterns for pairs of lasers that were used to calculate the fringe visibility are shown in the insets. The fringe visibility was above 0.9 for all pairs of lasers that belong to the same cluster and below 0.15 otherwise, serve as a clear verification for the existence of 4 distinct synchronized clusters.

The GCD of network loops is a global feature indicating that local changes in network connectivity, e.g. addition/deletion of couplings, can switch phase synchronization of remote nodes (Figs. 1 and 2). Such remote switching enables control on the number of clusters of synchronized lasers, which is highly desired for example in multi-user communication networks. Probably the simplest way to control multi-user synchronization is by means of a master laser node with an adjustable self-feedback loop that controls the number of clusters (Fig. 3a-d). The number of clusters, each denoted by a different colour, is determined by the GCD of the entire network loop and the self-feedback loop of the master laser node. We hypothesize, and validate in the next paragraph, that since optical phase information is transferred in accordance to the network connectivity, a $n \cdot \tau$ self-feedback loop is equivalent to a $n$ lasers loop, thus applying the GCD rule to the length of the feedback loop as if it were a loop of real lasers. A more compound remote switching scheme consists of a network with multiple loops, where $2^n$ different network configurations are formed using $n$ switches (Fig. 3e). For $n$=3, on/off switches $S_1$, $S_2$ and $S_3$ form three different loops of sizes $M$, $K$ and $L$ which control the number of clusters together with the backbone loop of size $N$.

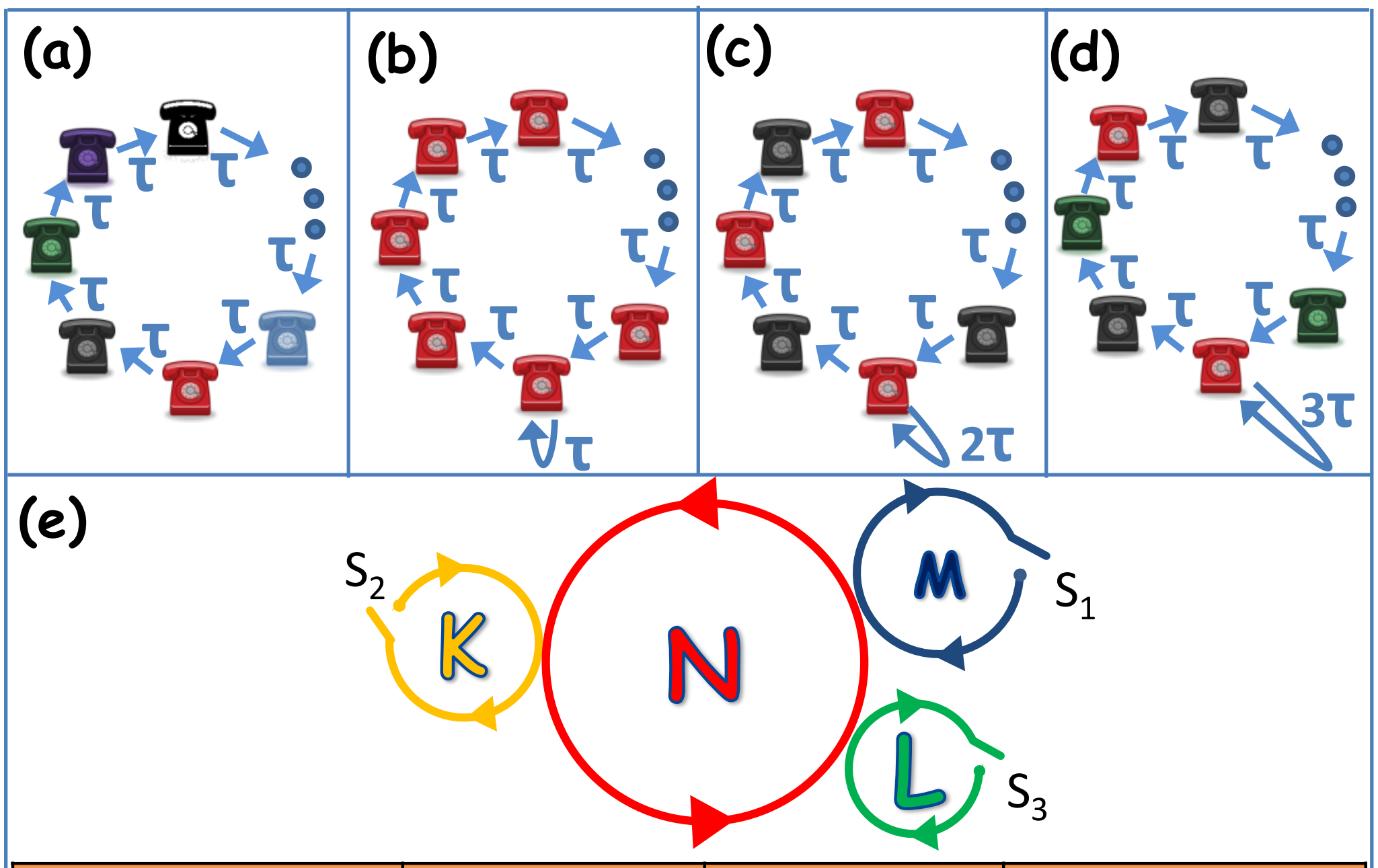

| $(s_1=0, s_2=0, s_3=0)$ | (0,1,0) | (1,1,0) | (1,1,1) |
|---|---|---|---|
| GCD(N) | GCD(N,K) | GCD(N,M,K) | GCD(N,M,K,L) |

**Figure 3| Remote switching to control multi-user synchronization in communication networks. a-d, Communication loop networks with time delay $\tau$, enabling multi-user communication controlled by a self-feedback loop to the "master" node. The number of distinct synchronized clusters is given by the GCD of the entire loop size and the master`s self-feedback loop. e, A multiple ring network of sizes M, K, L with $2^3$ possible different synchronized formations selected by the *3* On/Off switches, $S_1$, $S_2$ and $S_3$. The number of clusters is determined by the GCD of all the closed loops.**

The Mermin-Wagner theorem predicts that for low dimensional networks synchronization must eventually degrade as the network size increases[19,20]. The observed high fringe visibility values (>0.9) for 16 and 8 laser networks alike indicate that finite size effects may only apply to much larger networks, which require further investigations. Nevertheless we do expect that in high dimensional networks, where the Hamilton path among all nodes does not scale with the size of the network, the synchronization state is governed by the GCD rule.

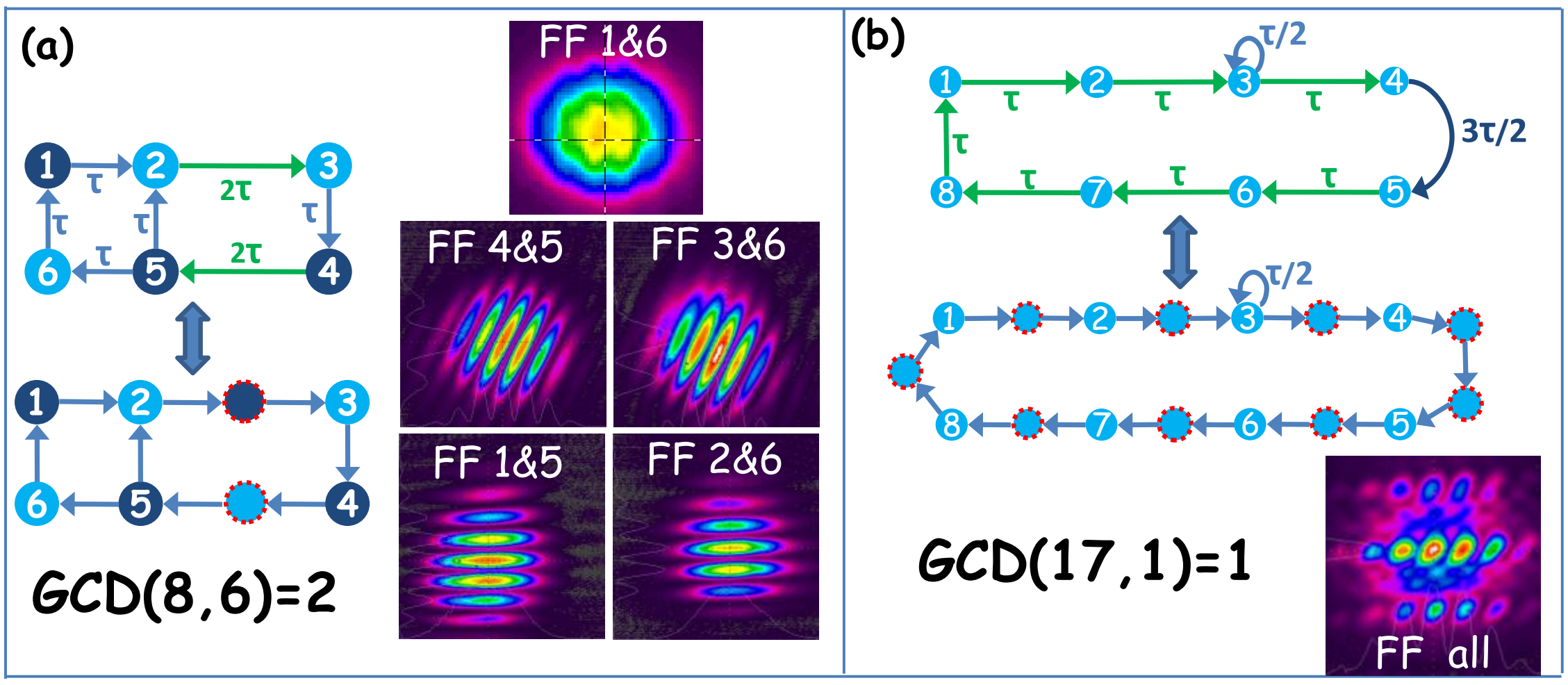

**Figure 4| Experimental phase synchronization of heterogeneous networks. The phase dynamics of networks with heterogeneous commensurate time-delays is determined by resorting to equivalent homogeneous networks where additional imaginary lasers split long delays to homogeneous short delay segments. a, A heterogeneous network of 6 lasers with $\tau$=4 ns and 2$\tau$=8 ns time delays (top) and the equivalent homogenous network of 8 lasers (bottom). The FF interference distributions of different pairs of lasers indicate the formation of two clusters GCD(8,6)=2. b, A heterogeneous network of 8 lasers with $\tau$/2=2 ns, $\tau$= 4 ns , and 3$\tau$/2=6 ns time delays (top) and the equivalent homogenous network of 17 lasers (bottom). The FF interference distribution of all lasers indicates a single synchronized cluster GCD(17,1)=1.**

The established GCD rule only applies to homogeneous networks, which are not suitable for communication networks that have heterogeneous distances among users. Thus, we extend the GCD rule to include the dynamics of heterogeneous networks with commensurate ratios among the delays [21]. This extension is achieved by resorting to equivalent homogeneous networks where imaginary lasers are added so as to split delays to homogeneously shorter delay segments. Then we apply the GCD rule to the equivalent homogeneous network to find the actual number of clusters in the heterogeneous network. We experimentally examined a heterogeneous network of 6 lasers with $\tau$ and $2\tau$ time delays (Fig. 4a, upper sketch). The equivalent homogenous network consists of two additional imaginary lasers so as to form 8 and 6 laser loops (Fig. 4a, lower sketch), leading to two synchronized clusters GCD(8,6)=2. The experimental results indeed revealed two synchronized clusters, lasers (1,4,5) and lasers (2,3,6), based on the contrast of the interference fringes in the FF patterns which is above 0.9 for any pair of lasers that belong to the same cluster and below 0.1 otherwise (Fig. 4a). The role of the GCD was further examined for a more compound heterogeneous network consisting of eight lasers and three different time delays, $\tau/2$, $\tau$ and $3\tau/2$ (Fig. 4b, upper sketch**).** The equivalent homogenous network with equal time delays of $\tau/2$ has additional nine imaginary lasers that form a directed loop of 17 lasers with an additional single self-feedback loop (Fig. 4b, lower sketch). This equivalent homogeneous network has one synchronized cluster GCD(17,1)=1, as confirmed by the high contrast fringes in the FF interference pattern of all the eight lasers (Fig. 4b).

Our results offer a route to achieve robust, reliable, broadband multi-user communication networks as well as the remote switching mechanism to control advanced multi-user protocols. The relation between network synchronization and number theory suggests that synchronization may also be used to perform hard combinatorial tasks in optical computers and highly versatile multiplexing tasks over shared communication lines. In addition, the GCD could play a role in intensity synchronization of chaotic lasers offering the prospect of an all optical secure communication. We numerically confirmed the role of GCD for phase synchronization of homogeneous/heterogeneous delay networks using the Kuramoto model that describes a general class of oscillators (Supplementary Information). Accordingly our approach and results could be applied to a variety of coupled oscillators in electrical, biological, chemical and climatic phenomena.

## METHODS SUMMARY

**Degenerate cavity.** The degenerate cavity (Fig. 1a) is comprised of a Nd-Yag crystal gain medium that can support several independent laser channels, front and rear output couplers (O.C.), a mask of an array of apertures of 0.2 mm diameters and 0.3 mm spacing that forms the different laser channels, and two lenses in a 4*f* telescope arrangement. The telescope images the mask plane to the front O.C. plane so as to ensure that different lasers do not interact in the gain medium, and thus remain uncoupled[9]. A calcite crystal placed inside the cavity displaces each beam into two parallel beams with ordinary (Ô) and extra-ordinary (Ê) polarization states, so that a mask with *N* apertures would lead to *2N* laser beams. From the front of the cavity these *2N* beams emerge spatially separated, while from the rear they emerge folded on to each other, as shown by the example of 16 laser beams in Fig. 1a.

**Coupling arrangement.** A variety of coupling connectivities can be realized by controlling the angular orientations of the four coupling mirrors R1, $R_2$, $R_3$ and $R_4$ in the coupling arrangement. Unidirectional time delayed coupling is achieved by means of a Faraday rotator positioned along the beams paths, a focusing lens *f* positioned at a distance *f* from the rear O.C., a polarizing beam splitter (PBS), and the four coupling mirrors all placed within the Rayleigh focal range of the focusing lens [9,22].

With a reflectivity of 40% for $R_2$ and $R_4$ and 100% for $R_1$ and $R_3$, four nearly equally intense mirror images of the transverse field $E(x,y)$ are reflected back towards the lasers. Each mirror image $E(-(x-x_0),-(y-y_0))$ can be reflected around a different point $(x_0,y_0)$ that denotes the self-reflecting point of the corresponding mirror and is determined by its angular orientation. The Faraday rotator, rotates the polarization state so as to couple Ô polarized lasers to Ê polarized lasers via mirrors $R_1$ or $R_2$ and Ê polarized lasers to the Ô polarized lasers via mirrors $R_3$ or $R_{4,}$ thereby allowing for unidirectional coupling. By independently controlling the angular orientations of all four coupling mirrors, we realized a variety of connectivities between the lasers, whereby each mirror connects pairs of lasers of orthogonal polarizations that are symmetric around its self-reflecting point. Additional coupling mirrors, non-polarizing beam splitters and lenses, (not shown in Fig. 1), were sometimes added in order to introduce delayed self-feedback and heterogeneous coupling delays (Supplementary Information).

**Detection arrangement.** The level of phase synchronization is quantified by the second order correlation the electric fields of the lasers which is directly measured by the contrast of the fringes in the FF intensity interference pattern detected by the CCD. A linear polarizer oriented at an angle of 45°, (not show in Fig. 1), was placed before the CCD in order to measure the interference of orthogonally polarized lasers. In general any cluster formations can be determined by interfering different pairs of lasers in the network, as can be seen in Figs. 1c, 2e and 4a. In some cases however, a single measurement of the FF interference pattern is sufficient to observe the cluster formation, as exemplified in the FF pattern of all 16 lasers appearing in the centre of the sketch on the right in Fig 1c. In this example, the directionality of the high contrast fringes indicates that synchronization occurs only between lasers positioned along the same vertical column of the lasers array, so two synchronized clusters are formed. Similarly, from the single measurement of the high contrast fringes along both horizontal and vertical directions in the FF interference pattern shown in Fig. 2b, one can deduce that synchronization occurs between all lasers in the network.

**Acknowledgements** The work was supported in part by the Israel–U.S.A. Binational Science Foundation and the ISF Bikura grant.

# Supplementary Information for: Controlling synchronization in large laser networks using number theory

Micha Nixon[1], Moti Fridman[1], Eitan Ronen[1], Asher A. Friesem[1], Nir Davidson[1] & Ido Kanter[2].

[1]*Dept. of Physics of Complex Systems, Weizmann Institute of Science, Rehovot 76100, Israel.*

[2]*Dept. of Physics, Bar-Ilan University, Ramat-Gan 52900, Israel.*

## The coupling configurations

Our coupling arrangement includes four mirrors located at the back focal plane of a focusing lens [1,2]. Each mirror inverts the incident light relative to its own self-reflecting point which is determined by the mirror orientation. To obtain unidirectional coupling from laser A to laser B, the light from laser A must be injected into laser B while the light from laser B must be directed elsewhere. This is achieved by utilizing the polarization degree of freedom of each laser. Specifically, if laser A has Ô polarization and laser B orthogonal Ê polarization, then since each pair of coupling mirrors is placed along a different arm of a polarizing beam splitter (PBS), each pair of mirrors interacts with either Ô or Ê polarized light. Thus, although both lasers A and B can spatially overlap and even emerge from the same hole in the mask, each will follow a different path though the coupling arrangement and will be reflected back from a different mirror. Thereby, enabling unidirectional coupling of light from laser A to B via a mirror that interacts with Ô polarized light while allowing the light emerging from laser B to be directed elsewhere.

The self-reflecting points of the mirrors that were used to couple the lasers in the eight laser network in Fig. (2**e**) of the manuscript, are presented as black stars in Figs. 1. They were located in the rear near-field (NF) image plane and are shown along with four coloured arrows that denote which pairs of lasers are coupled by which mirrors. Also shown is a schematic sketch of the rearranged laser locations using the same colour code.

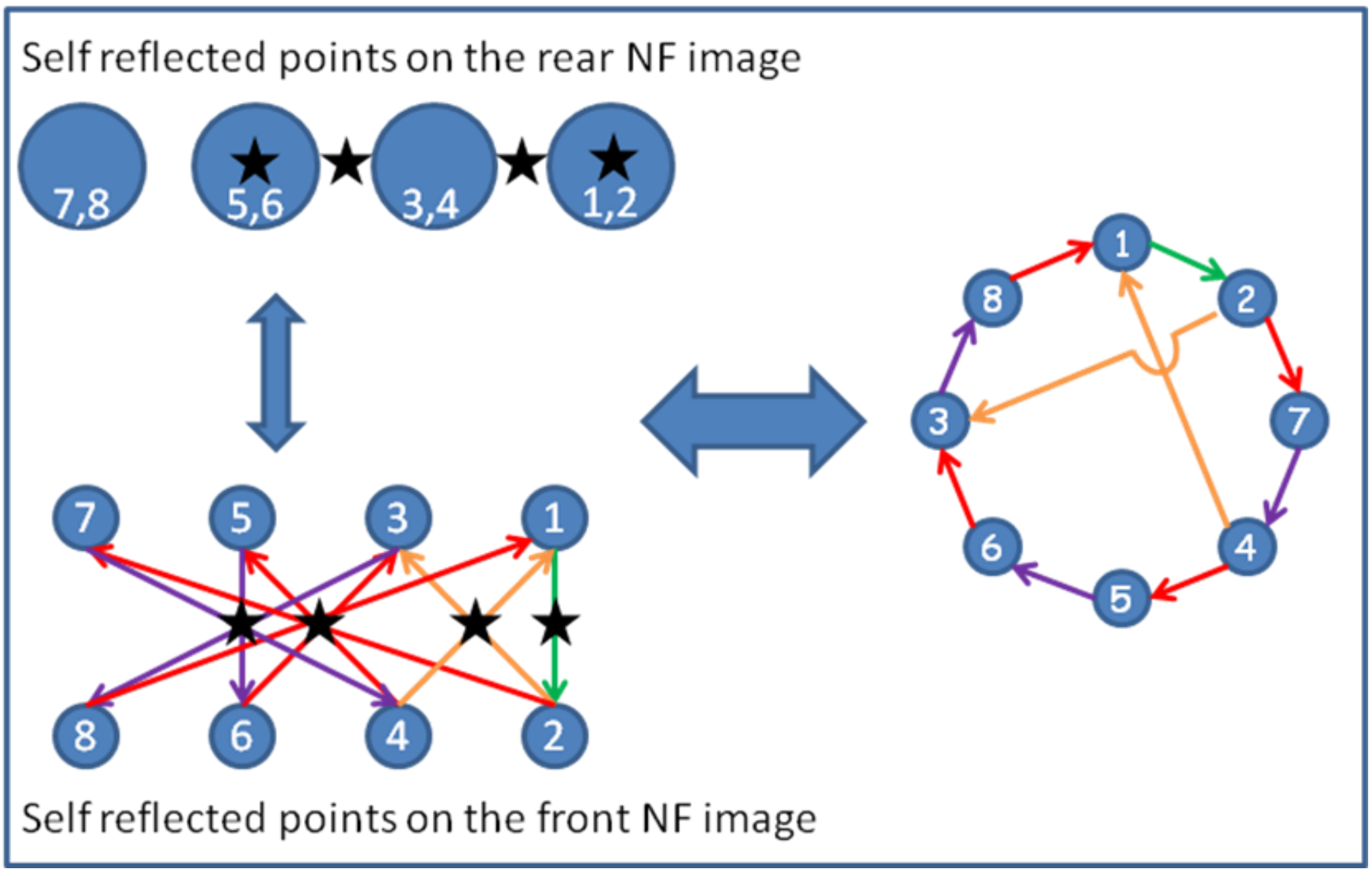

**Figs. 1 Schematic sketch for the lasers connectivity corresponding to the experimental configurations shown in Fig. 2e of the manuscript. Black stars denote the self-reflecting points of the coupling mirrors, different colour arrows denote the coupling induced by a different coupling mirror.**

Heterogeneous unidirectional coupling was introduced by adding two lenses placed in 4f telescope that images the coupling mirror plane to a farther away plain enabling coupling with a differs delay times. Supplementary figure 2 shows a coupling configuration that enables such heterogeneous unidirectional coupling to occur, as shown in the figure taking $f_1$=300 *mm* we obtain a total coupling distance of $4f_1$=1.2 *m* corresponding to a $\tau$=4 *ns* coupling delay time for mirror $R_1$, $R_2$ and $R_4$, for $f_2$=150 *mm* the additional telescope adds $8f_2$=1.2 *m* to the coupling distance resulting in a coupling delay time of $\tau$=8 *ns* for mirror $R_3$.

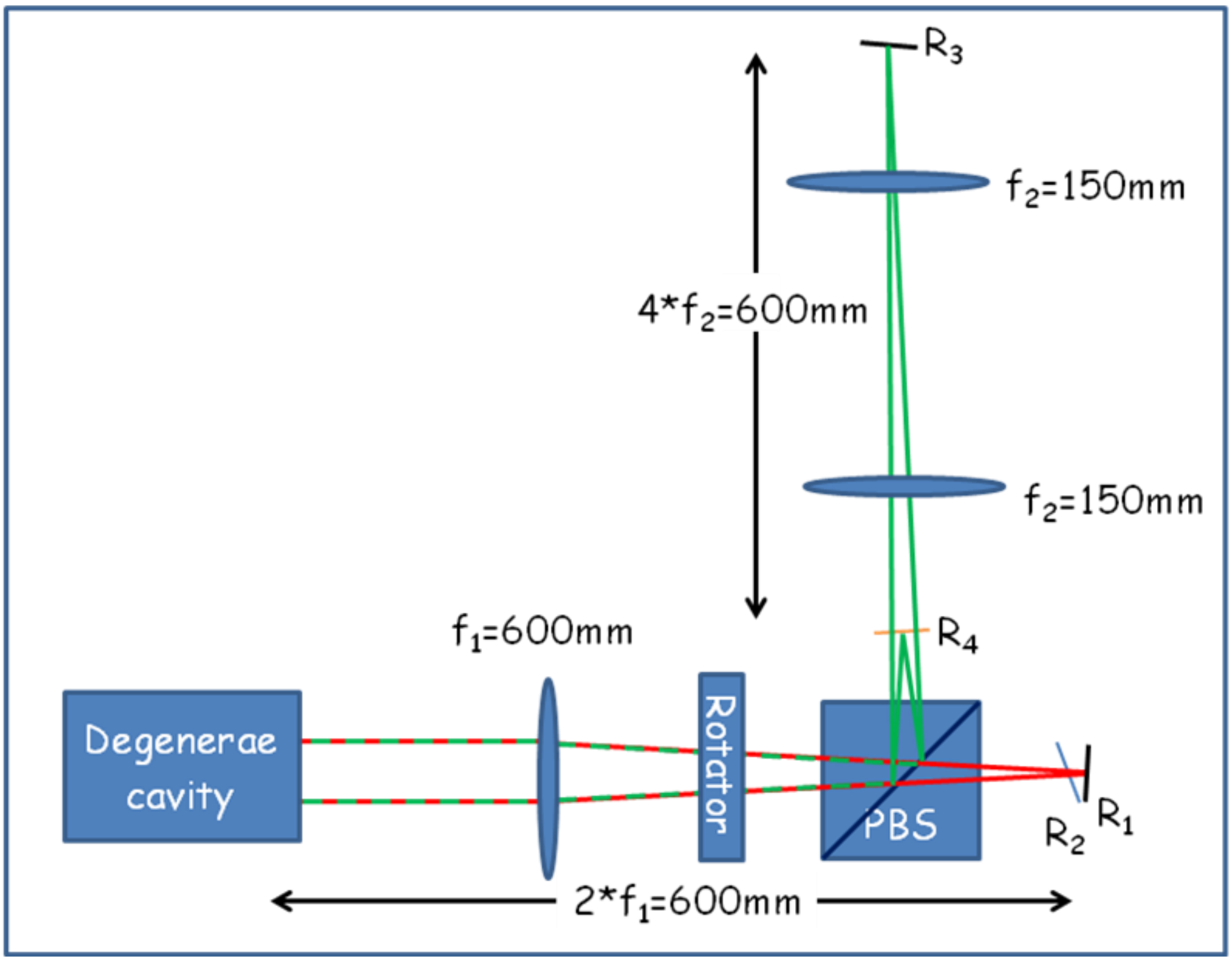

**Fig. 2 Arrangement for the coupling connectivity allowing for heterogeneous commensurate coupling delay times.**

Delayed self-feedback is introduced by adding a 40% partial reflector that directs part of the light towards a self-feedback coupling mirror, as shown schematically in Figs. 3. This light does not pass through the Faraday rotator so it maintains its polarization state. By adjusting the self reflecting point of the self-feedback coupling mirror to lie directly on top of a laser positioned at the edge of the array, light is only coupled from that laser back to itself. By adding the polarizer we can ensure that only a single laser is coupled to its own signal delayed by $\tau=4f_2/c$.

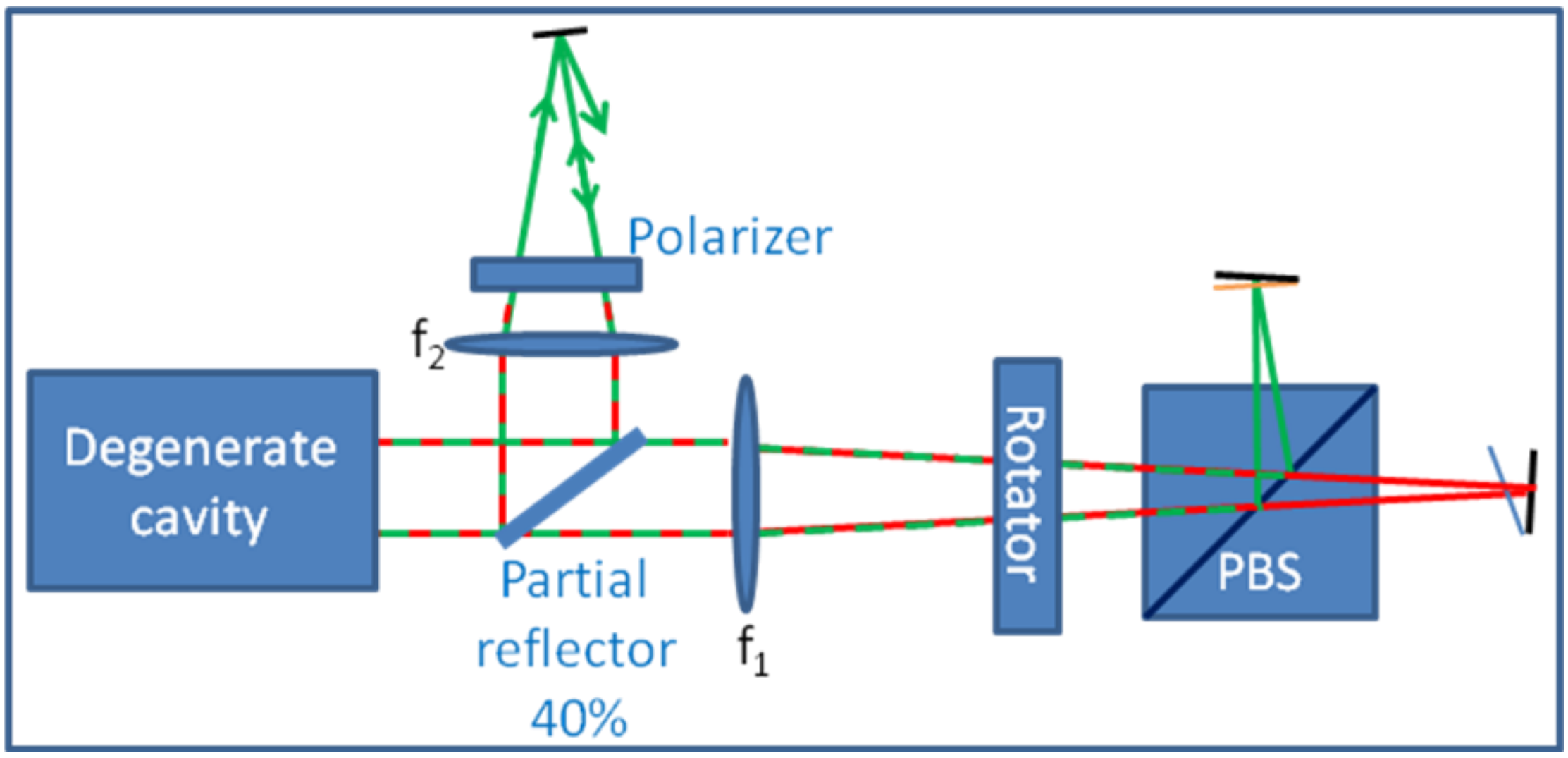

**Fig. 3 Arrangement for introducing delayed self feedback.**

## The Kuramoto model

To simulate the phase dynamics of our time-delayed coupled lasers, we used the Kuramoto model which describes the dynamics of coupled oscillators [3]. The phases of the individual oscillators $\varphi_i$, are obtained by solving

$$\frac{d\varphi_i}{dt} = \omega_i + \sum_j \sigma_{i,j} \sin\left(\varphi_j(t-\tau) - \varphi_i(t)\right),$$

where $\sigma_{i,j}$ is the coupling matrix, $\tau$ is the coupling delay time, $\omega_i$ is the detuning of oscillator *i* from the mean frequency. It should be noted that in our experiments the coherence time depends on the spectral bandwidth of our lasers, whereas the Kuramoto model deals with the dynamics of single mode oscillator. As a result, we artificially added a finite coherence time to the Kuramoto model in the form of phase noise [1,4]. Specifically, we solved the equation when the frequency detuning $\omega_i$ of the oscillators changes stepwise in time with a random and uncorrelated Gaussian distribution of zero mean value and a width of$\omega_0$. The duration of the time steps determine the coherence time of the individual oscillators.

The level of synchronization between two oscillators was quantified by calculating the second order correlation function of the phases. This second order phase correlation function mathematically defines the fringe visibility that is measure for lasers having the same constant amplitudes (1,4,5).

We simulated the Kuramoto model for a wide verity of scenarios similar to the experimental ones and found that for all cases the calculated results are in good agreement with the experimental results. Thus suggests that the GCD rules presented in the manuscript describe a more general class of coupled oscillators and not just lasers.